# Translating Common Security Assertions Across Processor Designs: A RISC-V Case Study

Sharjeel Imtiaz, Uljana Reinsalu and Tara Ghasempouri
Department of Computer Systems, Tallinn University of Technology, Tallinn, Estonia
{sharjeel.imtiaz, uljana.reinsalu, tara.ghasempouri}@taltech.ee

*Abstract*—RISC-V is gaining popularity for its adaptability and cost-effectiveness in processor design. With the increasing adoption of RISC-V, the importance of implementing robust security verification has grown significantly. In the state of the art, various approaches have been developed to strengthen the security verification process. Among these methods, assertion-based security verification has proven to be a promising approach for ensuring that security features are effectively met. To this end, some approaches manually define security assertions for processor designs; however, these manual methods require significant time, cost, and human expertise. Consequently, recent approaches focus on translating pre-defined security assertions from one design to another. Nonetheless, these methods are not primarily centered on processor security, particularly RISC-V. Furthermore, many of these approaches have not been validated against real-world attacks, such as hardware Trojans. In this work, we introduce a methodology for translating security assertions across processors with different architectures, using RISC-V as a case study. Our approach reduces time and cost compared to developing security assertions manually from the outset. Our methodology was applied to five critical security modules with assertion translation achieving nearly 100% success across all modules. These results validate the efficacy of our approach and highlight its potential for enhancing security verification in modern processor designs. The effectiveness of the translated assertions was rigorously tested against hardware Trojans defined by large language models (LLMs), demonstrating their reliability in detecting security breaches.

## I. Introduction

RISC-V, an open-source instruction set architecture(ISA) [1–3], is popular for its modular design, allowing customization without licensing restrictions. Its flexibility makes it valuable for both industry and academia [4, 5], fostering global collaboration and innovation in processor development. As RISC-V expands, ensuring its security has become a major focus. In this regard, security verification [6–8] is crucial for maintaining the integrity of modern processors such as RISC-V, as its open-source nature poses unique challenges. With the increasing complexity of processor designs, verification techniques now address both functional correctness and security assurance. Semi-formal methods [9, 10], including assertion-based verification [11–14], have become vital in detecting vulnerabilities such as hardware Trojans [15]. Assertion-based verification allows early detection of security issues during the design phase, making it essential for protecting processors. Given the open-source and modular architecture of RISC-V, complete security verification [16] is critical to its future development.

Manual assertion-based security verification is a critical but resource-intensive process to ensure processor security. This approach thoroughly reviews processor specifications and security requirements, then manually writes assertions to check security properties such as access control [17] and information flow. Furthermore, manual security verification processes also extend to the verification of hardware and firmware interactions [18], a crucial aspect of system-on-chip (SoC) security. Engineers must understand the architecture in detail [19–21], which includes studying the processor design, reading relevant documentation, and identifying potential security risks. After creating the assertions, they run simulations to test for vulnerabilities, which can be a time-consuming process that requires iteration and fine-tuning.

In this regard, security assertion translation plays a crucial role in ensuring that hardware designs maintain their intended security requirements, even when adapted across different architectures. This technique allows security assertions, which are typically written to verify security-critical functions in one processor, to be reused in other processor designs, reducing the need to rewrite assertions from scratch. Automated translation frameworks streamline this process, adjusting signal variable names, multi-layer signals, and logical conditions to ensure that assertions are compatible with the new hardware without losing their intended security functions. One notable tool in this area is Transys [22], which translates security-critical properties between hardware designs by adjusting logical preconditions and other constraints.

Thus all in all manual generation of security properties for processors, as noted in [19], demands substantial time, effort, and specialized skills, making it costly. The challenge grows with modern processor complexity. On the other hand, translating security assertions between hardware designs offers a solution. Tools like Transys [22] assist with the process of security assertion translation, but they lack support for RISC-V-specific architecture. However, both manual and automated approaches share a major flaw as neither has been rigorously tested against real-world threats like hardware Trojans. Our work aims to fill the above gaps in processor security verification.

Our work addresses the challenges of manual security assertion definition and automated translation, focusing on enhancing RISC-V processors and testing them with realistic threats like hardware Trojans. We developed a method that reduces the cost and time of the security verification process by translating security assertions from one architecture to another. By targeting high-risk modules within the RISC-V architecture, based on prior research [23, 24], we adapted our approach for broader use. Unlike previous efforts, we rigorously tested translated assertions against hardware Trojans [25] generated by LLMs(ChatGPT) [26], ensuring robust detection of security breaches in real-world scenarios.

The contributions of this paper are listed as follows:

- We created a methodology for translating security assertions, thus assertions can be reused across different processor architectures, using RISC-V as a case study. This approach reduces the time and cost typically involved in developing security assertions from scratch for similar architectures.
- The translated security assertions were rigorously tested against hardware Trojans generated by large language models (LLM), demonstrating their reliability in detecting security breaches.
- We introduced security metrics to assess and quantify the processor's resilience against hardware Trojans, providing a consistent framework for security evaluation.

The rest of this paper is structured as follows. The preliminary data are explained in section II and the proposed methodology is presented in section III. Section IV presents the experimental results, and section V concludes the paper.

## II. Preliminaries

In this section, we briefly explain the definitions and concepts used in this paper.

This work was supported by the Estonian Research Council grants PSG837.

*Definition 1:* A **security property** in this study refers to a critical aspect of the processor's design, where overlooking it could lead to potential vulnerabilities. These properties are typically specified in the processor's documentation [1, 2].

*Definition 2:* A **security assertion** is a logical expression that must hold true during the operation of a design. In this study, assertions follow the structure `assert (antecedent |=> consequent)`, meaning that whenever the antecedent condition is met, the consequent must follow [27]. These assertions are used to verify that the actual implementation aligns with the behaviors defined by the `security properties`, ensuring the system responds correctly when facing vulnerabilities or hardware Trojan attacks.

*Definition 3:* **Temporal pattern ##N** in terms of SystemVerilog assertions (SVA), where the ##N construct is used to indicate a wait for N clock cycles [27]. Here's the example in SystemVerilog terms: `assert property (antecedent |-> ##N consequent);` This indicates that whenever the antecedent occurs after N clock cycles, the consequent should occurs.

*Definition 4:* **The temporal pattern $past** is used to refer to a condition that must have held true at a previous time step [27]. The syntax can be expressed as: `assert property ($past(condition));` This represents the property or expression that you want to evaluate in the past. For example, to assert that a signal signal_x was true two cycles ago: `assert property (signal_x |-> ##2 $past(signal_x));`

*Definition 5:* **NS31A** [28] is a 32-bit RISC-V CPU core with a single-issue, in-order 4-stage pipeline. It supports the RV32IMAF instruction set and is designed for functional safety in automotive applications, compliant with ISO 26262 ASIL D.

*Definition 6:* **Ibex** [29] is a 32-bit open-source RISC-V 2 stage pipeline CPU core designed for embedded control applications. Written in SystemVerilog, it supports various RISC-V extensions, including Integer (I/E), Multiplication and Division (M), Compressed (C), and Bit Manipulation (B).

*Definition 7:* **Direct mapped signal** refers to the process of connecting one signal directly to another without any intermediate logic or processing. An example of direct signal mapping in Register Transfer Level (RTL) would be as follows:

```
assign Internal_signal_1 = Top_module_signal_1
                         & Internal_signal_2;
```

*Definition 8:* **Indirect mapped signal** refers to connecting signals through one or more intermediate signals or logic elements. In this approach, a signal is not directly linked to another but relies on intermediary signals for connection i.e. given RTL example:

```
assign Internal_signal_2 = Top_module_signal_2
                         & Top_module_signal_3;
```

In this case, `Internal_signal_2` is directly mapped signal (Defination7) to `Top_module_signal_2` and `Top_module_signal_3`. On the other hand, `Internal_signal_1` (Definition 7) is defined as :

```
assign Internal_signal_1 = Top_module_signal_1
                         & Internal_signal_2;
```

is indirectly connected to `Top_module_signal_2` and `Top_module_signal_3` through `Internal_signal_2`.

## III. PROPOSED METHODOLOGY

The overview of the proposed methodology is presented in Figure 1, which divides the process into three key steps. As can be seen, this methodology begins by selecting a module under security verification [23, 24] of the RISC-V processor, Key modules identified include Physical Memory Protection (PMP), Control Status Registers (CSR), Control Flow, Debug Operation, and Exception, Interrupt, and Trap Handling each crucial for maintaining system security. These modules are responsible for managing functions that, if compromised, could expose vulnerabilities, such as protecting memory access and securing program execution paths. By contrast, other modules like the ALU and MUX, which handle computation, pose lower security risks. Once identified, the modules first undergo security verification. In Step 1, security assertions are translated from one RISC-V processor to another, ensuring compatibility and effectiveness within the different processor designs of the same architecture. In Step 2, hardware Trojans are injected to test the robustness of these assertions against simulated security threats. Step 3 involves measuring the effectiveness of the translated assertions using two proposed metrics to evaluate Trojan detection.

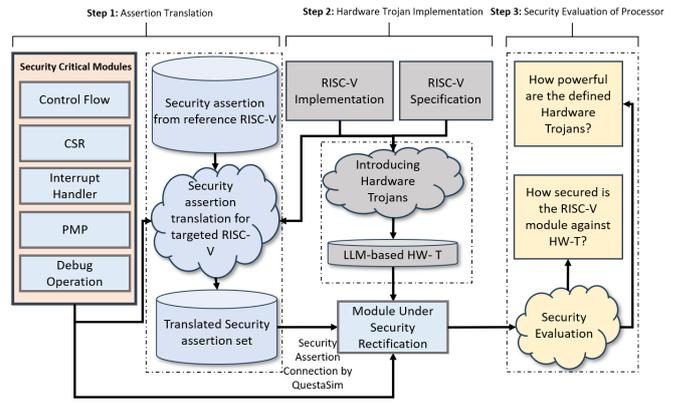

Figure 1: Overview of the methodology for security assertion translation and evaluation

We utilize semi-formal verification techniques, incorporating assertions to develop a security verification tool tailored for processor architectures. This tool systematically checks for security vulnerabilities by ensuring that predefined conditions hold throughout the processor's execution. The details of each step are described as follows.

### A. Assertion Translation (Step 1):

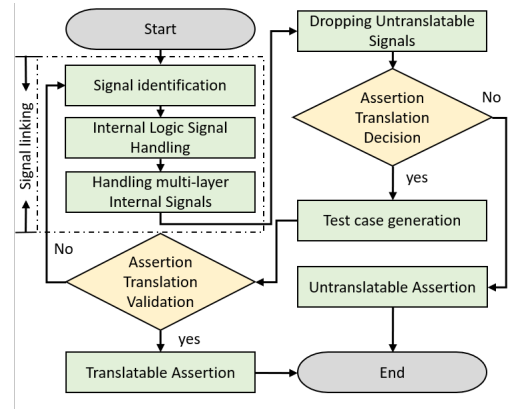

Figure 2: Flowchart illustrating the process of assertion translation.

The proposed methodology begins by selecting modules under security verification within the RISC-V architecture. In step 1 of assertion translation, we gather security assertions from state-of-the-art processors [19] and adapt them for target RISC-V architecture. This involves mapping signals from the original datasets to corresponding signals in the target RISC-V processor to ensure the assertions align with modules under security verification. In this security assertion translation, we predominantly use temporal security assertions, such as the temporal patterns ##N (Definition3) and $past (Definition4).

Figure 2 The flow begins with the *Signal Linking process*, which consists of three main stages. First, in *Signal Identification*, signals from the original assertion are analyzed to confirm their presence

in both the RTL logic and the module under security verification. Next, during *Internal Logic Signal Handling*, directly mapped signals (Definition7) are managed for instance, in the RTL code example:

Listing 1: Code snippet intrupt request handle & enable.
```
assign handle_irq = ~debug_mode_q & ~debug_single_step_i
                    & ~nmi_mode_q & (irq_nm |
                    (irq_pending_i & irq_enabled));
assign irq_enabled = csr_mstatus_mie_i
                    | (priv_mode_i == PRIV_LVL_U);
```

A signal like `handle_irq` is directly linked to internal signals such as `irq_enabled` and `irq_nm`. Finally, in *Handling Multi-layer Internal Signals*, indirect Mapped signals (Definitions 8), like `csr_mstatus_mie_i` and `priv_mode_i`, which influence `irq_enabled`, are traced to ensure comprehensive signal handling. In the *Dropping Untranslatable Signals stage*, an assessment is made to determine whether any signals from the original assertion are irrelevant to the RTL logic of the selected processor architecture or not present in the RTL, leading to the removal of those that do not align. Additionally, any extra RTL signals unrelated to the original assertion are eliminated. In the *Test Case generation* step, test cases are created based on the translated assertions, with further adjustments made to the security assertions as necessary. The *Untranslatable Assertion* stage indicates that the assertion cannot be translated due to reasons such as the absence of exact signals or mismatched behavior in the selected processor RTL. Finally, in the *Translatable Assertion* step, confirmation is provided that the translated security assertion has been tested and successfully passed simulation.

signal is `csr_op_i`, which indicates the write operation in the Ibex(Definition6) CSR. These signals control write enable and address operations within the CSR module. Moving to Step 2 *Internal Logic Signal Handling*, the internal logic signals are further systematically examined. `csr_we_int` and `mstatus_en` are mapped from the top-level CSR module, ensuring that their roles are correctly interpreted. In Step 3 *Handling Multi-layer Internal Signals*, multi-layer dependencies are addressed. For example, `csr_we_int` is influenced by signals like `csr_wr`, `csr_op_en_i`, and `illegal_csr_insn_o`, which are essential for controlling write operations within the module. If any signals from the original assertion are not found in the Ibex(Definition6) core, they are eliminated in Step 4 *Dropping Untranslatable Signals*. In this case, however, all signals are translatable and correspond to their intended roles. Finally, in Step 5 *Test Case generation*, the translated assertion is integrated into the RTL design, and a test case is developed to ensure its functionality. The final assertion checks that the MSTATUS register is not written unless the correct address is provided, with the structure refined for optimal performance within the Ibex(Definition6) core. The final assertion is structured as follows:

Listing 3: Translated security assertion for IBEX CSR module
```
property csr_write_with_matchaddr;
@(posedge clk_i) (csr_addr_i != CSR_MSTATUS)
      && (csr_op_i != CSR_OP_WRITE)
      |-> (csr_we_int==0 && mstatus_en==0);
end property
CSR_2: assert property (csr_write_with_matchaddr) else
$error("Test_failed_for_mstatus_write");}
```

### B. Hardware Trojan development (Step 2):

Hardware Trojan development using Large Language Models (LLMs) begins by uploading several RTL files, including the processor module under security verification and the translated assertion RTL file, into ChatGPT-4 for processing. The LLM analyzes these files to identify vulnerabilities and potential weak points in the design based on the given prompt, streamlining the development of the Trojan.

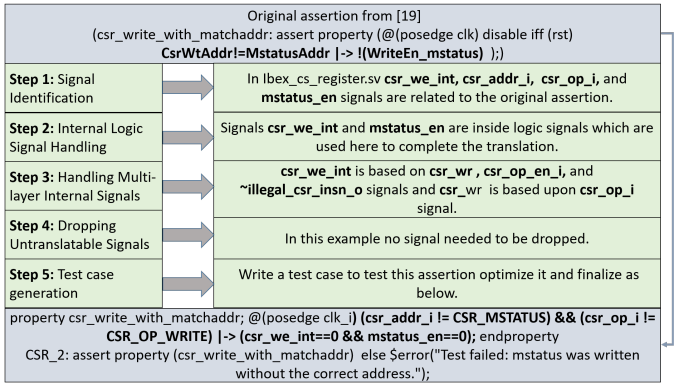

Figure 3: Example of translating an assertion for the Ibex [29] CSR module.

The example in Figure 3 demonstrates the translation of an assertion from a dataset for the Control Status Register (CSR) module within RISC-V processors. The original assertion from the NS31A(Definition6) RISC-V processor is expressed as follows:

Listing 2: Orignal secuirty assertion from dataset
```
assert property (@(posedge clk) disable iff (rst)
      CsrWtAddr != MstatusAddr|-> !(WriteEn_mstatus));
```

This security assertion verifies that CSR writes are only permitted when the addresses match, specifically targeting the MSTATUS address. The translation process of security assertions begins with Step 1, *Signal identification*, where relevant signals from the original design are identified. For example, an assertion checking conditions involving the MSTATUS address requires identifying signals such as `csr_we_int` and `csr_addr_i` (corresponding to `CsrWtAddr` in the original assertion), as well as `mstatus_en` (matching `WriteEn_mstatus` in the original assertion) within the Ibex(Definition6) core. Additionally, there are extra signals in the Ibex(Definition6) core's CSR module that are crucial for completing the RTL logic, as the translated assertion would not function correctly without them. One such

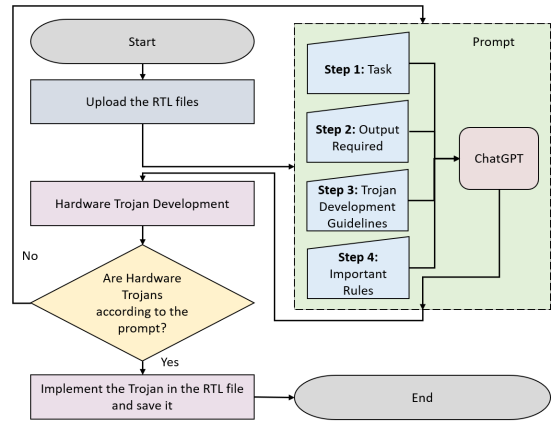

Figure 4: Flowchart for hardware Trojan generation using LLM(ChatGPT)

The process of hardware Trojan development using ChatGPT is illustrated in the flowchart shown in Figure 4 the flowchart. It begins with *Uploading the RTL files* of the module under security verification, such as the PMP of a RISC-V processor. These files are then loaded into the interface of LLM(ChatGPT) for processing. The next step involves crafting a detailed *Prompt* for ChatGPT, divided into four parts: *Task, Output Required, Trojan Development Guidelines*, and *Important Rules*. In the *Task* section, the prompt instructs ChatGPT to generate Trojans according to the translated assertions of the selected module under security verification, focusing on areas where these assertions are applied. In the *Output*

*Required* section, it specifies the generation of two SystemVerilog files: one with the embedded Trojans in the RTL and another with test cases designed to activate the Trojans, which can later be verified through simulation. In *Trojan Development Guidelines*, the prompt directs that Trojans should trigger based on specific signals, including certain bits of multi-bit signals, with the payload affecting relevant signals to alter the logic within the selected RTL area. The *Important Rules* strictly instruct ChatGPT not to modify the overall RTL design; only the Trojan should be embedded, with no changes to modules or signals, and if the module is combinational (e.g., PMP), no clock or reset signals should be added. Following this, the hardware Trojans are developed and implemented into the RTL file if they align with the prompt.

## C. Security Evaluation of Process (Step 3):

The Security Evaluation phase is important for assessing assertion translation effectiveness, the triggering probability of hardware Trojans, and the overall security of the RISC-V module.

Starting metric **How powerful are the defined Hardware Trojans?**, the strength of each Trojan is evaluated by calculating its Triggering probability. This involves analyzing the signals triggering the Trojan and the combinations of bits involved. The resulting probabilities, ranging from $10^{-1}$ to $10^{-22}$, show the Trojans' power, as they require rare but precise conditions to activate. Additionally, we assess whether the Trojan affects a single signal or disrupts the system's overall functionality. The significance of the disrupted logic is examined to identify weak points. To better quantify Trojan strength, the "Trojan Power Index (TPI)" is proposed.

$$\text{TPI} = \log_{10}\left(\frac{1}{P}\right) \quad (1)$$

Where: P represents the Trojan's triggering probability, determined by one combination out of the total possible combinations based on the signals and the total number of bits involved. For example, if three 1-bit signals are used, there are $2^3 = 8$ possible combinations, and the triggering probability $(1/2)^3 = 1.25 \times 10^{-1}$.

The second metric, **How secure is the RISC-V module against HW-T?**, the effectiveness of the translated assertions in detecting Hardware Trojans (HW-T) is evaluated. If all Trojans are intercepted, the assertion translation is considered effective, indicating strong protection for the security module. If any Trojans are missed, it highlights weaknesses in the assertion set, requiring a review of both Trojan generation and assertion translation. After adjustments, simulations are rerun for full coverage. To measure detection efficiency, the "Trojan Detection Efficiency Ratio (TDER)" metric is proposed.

$$\text{TDER} = \left(\frac{\text{Number of Trojans Detected by Translated Assertions}}{\text{Number of Trojans Generated}}\right) \times 100 \quad (2)$$

## IV. EXPERIMENTAL RESULTS

Experiments were conducted using Mentor Graphics QuestaSim with RTL files for the Ibex(Definition6) RISC-V processor sourced from GitHub [29]. A new project was set up in QuestaSim, compiling the necessary RTL files to monitor code correctness and assertion behavior. Debugging was done using QuestaSim's waveform tab to analyze errors in real time.

Table I: Security Assertion Translation & Trojan Detection

| Module Name | Assertions NS31A [19] | Translated Assertions | Translation % | Generated Trojans | Detected Trojans | Trojan Detection % |
|---|---|---|---|---|---|---|
| PMP | 7 | 7 | 100% | 7 | 7 | 100% |
| CSR | 7 | 7 | 100% | 7 | 7 | 100% |
| DO | 4 | 4 | 100% | 4 | 4 | 100% |
| ETI | 6 | 6 | 100% | 6 | 6 | 100% |
| CF | 9 | 9 | 100% | 9 | 9 | 100% |

Table I shows the translation and effectiveness of security assertions from NS31A [19](Definition5) to Ibex [29](Definition6) RISC-V modules. The *Module Name* column lists the modules (PMP, CSR, DO, ETI, CF) under security verification. *Assertion NS31A* indicates the number of assertions obtained from the NS31A processor for each module (e.g., PMP and CSR each have 7 assertions). *Translated Assertions* shows how many of these were successfully applied to the Ibex core, which is 100% for all modules, as shown in *Translation %*. The *Generated Trojan* column reflects the number of Trojans generated for each module, while *Detected Trojan* shows how many of these were successfully detected using the translated security assertions. Finally, *Trojan Detection %* from translated security assertions confirms a 100% Trojan detection rate for all modules under security verification.

Table II: Probability of triggering the Trojans in modules

| HW-T No. | Module Name | Triggering Probability | TPI | HW-T No. | Module Name | Triggering Probability | TPI |
|---|---|---|---|---|---|---|---|
| 1 | PMP | $1.91 \times 10^{-6}$ | 5.72 | 18 | DO | $1.2 \times 10^{-1}$ | 0.92 |
| 2 | PMP | $3.125 \times 10^{-2}$ | 1.50 | 19 | ETI | $7.8125 \times 10^{-3}$ | 2.11 |
| 3 | PMP | $2.117 \times 10^{-22}$ | 21.67 | 20 | ETI | $1.5625 \times 10^{-2}$ | 1.81 |
| 4 | PMP | $2.45 \times 10^{-4}$ | 3.61 | 21 | ETI | $1.56 \times 10^{-2}$ | 1.81 |
| 5 | PMP | $1.421 \times 10^{-14}$ | 13.85 | 22 | ETI | $1.907 \times 10^{-6}$ | 5.72 |
| 6 | PMP | $3.7 \times 10^{-9}$ | 8.43 | 23 | ETI | $1.2 \times 10^{-1}$ | 0.92 |
| 7 | PMP | $3.725 \times 10^{-9}$ | 8.43 | 24 | ETI | $1.22 \times 10^{-4}$ | 3.91 |
| 8 | CSR | $1.25 \times 10^{-1}$ | 0.90 | 25 | CF | $1.455 \times 10^{-11}$ | 10.84 |
| 9 | CSR | $2.5 \times 10^{-1}$ | 0.60 | 26 | CF | $2.91038 \times 10^{-11}$ | 10.54 |
| 10 | CSR | $2.44 \times 10^{-4}$ | 3.61 | 27 | CF | $3.9 \times 10^{-3}$ | 2.41 |
| 11 | CSR | $\& 1.525 \times 10^{-5}$ | 4.82 | 28 | CF | $3.9 \times 10^{-3}$ | 2.41 |
| 12 | CSR | $3.9 \times 10^{-5}$ | 4.41 | 29 | CF | $1.45 \times 10^{-11}$ | 10.84 |
| 13 | CSR | $2.44 \times 10^{-4}$ | 3.61 | 30 | CF | $2.91 \times 10^{-11}$ | 10.54 |
| 14 | CSR | $1.164 \times 10^{-10}$ | 9.94 | 31 | CF | $1.25 \times 10^{-1}$ | 0.90 |
| 15 | DO | $3.125 \times 10^{-2}$ | 1.50 | 32 | CF | $2.5 \times 10^{-1}$ | 0.60 |
| 16 | DO | $6.25 \times 10^{-2}$ | 1.20 | 33 | CF | $2.5 \times 10^{-1}$ | 0.60 |
| 17 | DO | $2.5 \times 10^{-1}$ | 0.60 | | | | |

Table II summarizes the triggering probabilities of hardware Trojans across different modules during security verification. Each Trojan, identified in the *HW-T No.* column, is implemented in modules like PMP, CSR, DO, ETI, and CF. The *Triggering Probability* column lists the likelihood of activation in scientific notation, with lower values indicating less chance of activation. The *TPI (Trojan Probability Index)* measures the difficulty of triggering each Trojan; higher values signify greater difficulty. In the PMP module, the probabilities range from $3.12 \times 10^{-2}$ to $2.117 \times 10^{-22}$. For example, HW-T3 has a maximum TPI of 21.67, indicating high difficulty, while HW-T2 is easier to trigger with a TPI of 1.50. In the CSR module, probabilities vary from $1.25 \times 10^{-1}$ to $1.164 \times 10^{-10}$, with TPI ranging from 0.60 to 9.94, highlighting HW-T8 and HW-T9 as more likely to be triggered. The DO module shows probabilities from $1.2 \times 10^{-1}$ to $6.25 \times 10^{-2}$ and TPI from 0.60 to 1.50, with HW-T17 and HW-T18 being particularly trigger-pron. For the ETI module, TPI ranges from 0.92 to 5.72, with probabilities between $1.2 \times 10^{-1}$ and $6.25 \times 10^{-6}$. While most Trojans are easy to trigger (e.g., HW-T23), HW-T22 is more challenging. The CF module has the broadest range of Trojans, with probabilities from $2.5 \times 10^{-1}$ to $2.91 \times 10^{-11}$ and TPI from 0.60 to 10.84. Notably, HW-T32 and HW-T33 show higher chances of being triggered.

## V. CONCLUSION

In conclusion, this research introduces a novel methodology to translate security assertions between processor architectures, exemplified by RISC-V. This approach reduces the time, cost, and complexity of developing security assertions from scratch for security verification of another processor architecture. By targeting modules under security verification, translated assertions successfully detected hardware Trojans developed using advanced AI techniques, achieving 100% detection success across all modules in simulations conducted in the QuestaSim environment.